\documentclass[a4paper, amsfonts, amssymb, amsmath, reprint, showkeys,prb, nofootinbib, twoside,superscriptaddress,citeautoscript,longbibliography]{revtex4-1}
\usepackage[english]{babel}
\usepackage[utf8]{inputenc}

\usepackage[colorinlistoftodos, color=green!40, prependcaption]{todonotes}
\usepackage{amsthm}
\usepackage{mathtools}
\usepackage{physics}
\usepackage{xcolor}
\usepackage{graphicx}
\usepackage[left=16mm,right=16mm,top=20mm,bottom=22mm,columnsep=15pt]{geometry} 
\usepackage{adjustbox}
\usepackage{placeins}
\usepackage[T1]{fontenc}
\usepackage{lipsum}
\usepackage{csquotes}

\usepackage[pdftex, pdftitle={Article}, pdfauthor={Author}]{hyperref} 
\usepackage{float}
\newcommand{\kB}{k_{\mathrm{B}}}
 
\setcitestyle{super}

\newcommand{\revision}{\textcolor{black} } 

\begin{document}
\title{Peptide Isomerization is Suppressed at the Air-Water Interface}
\author{Aditya N. Singh}
\affiliation{Department of Chemistry, University of California, Berkeley, CA 94720, USA}
\affiliation{Chemical Sciences Division, Lawrence Berkeley National Laboratory, Berkeley, CA 94720, USA}

\author{David T. Limmer}
\email[Corresponding author: ]{dlimmer@berkeley.edu}
\affiliation{Department of Chemistry, University of California, Berkeley, CA 94720, USA}
\affiliation{Chemical Sciences Division, Lawrence Berkeley National Laboratory, Berkeley, CA 94720, USA}
\affiliation{Materials Science Division, Lawrence Berkeley National Laboratory, Berkeley, CA 94720, USA}
\affiliation{Kavli Energy Nanoscience Institute at Berkeley, Berkeley, CA 94720, USA}

\date{\today} 

\begin{abstract}
We use molecular dynamics simulations to study the thermodynamics and kinetics of alanine dipeptide isomerization at the air-water interface. Thermodynamically, we find an affinity of the dipeptide to the interface. This affinity arises from stabilizing intramolecular interactions that become unshielded as the dipeptide is desolvated. Kinetically, we consider the rate of transitions between the $\alpha_L$ and $\beta$ conformations of alanine dipeptide and evaluate it as a continuous function of the distance from the interface using a recent extension of transition path sampling, TPS+U. The rate of isomerization at the Gibbs dividing surface is suppressed relative to the bulk by a factor of 3. Examination of the ensemble of transition states elucidates the role of solvent degrees of freedom in mediating favorable intramolecular interactions along the reaction pathway of isomerization. Near the air-water interface, water is less effective at mediating these intramolecular interactions. 
\end{abstract}

\maketitle

A number of chemical and biological processes have been purported to be catalyzed by air-water interfaces.\cite{Griffith.2012,Vaida.2017,Nam.2017,Morasch.2019,yadav2020chemistry,ruiz2020molecular,guo2021non,lee2019micrometer,stroberg2018cellular,nam2018abiotic}
Evidence of peptide-bond formation at the air-water interface has been reported using infrared reflection
spectroscopy\cite{Griffith.2012} and phosphorylation reactions have been claimed to become thermodynamically spontaneous in microdroplets generated with electrospray ionization.\cite{Nam.2017,nam2018abiotic} The interface of heated gas microbubbles has been shown to accumulate a range of biomolecules and act as a site for their condensation, enrichment and assembly.\cite{Morasch.2019} The intrinsic physical and chemical properties of the liquid-vapor interface have been implicated in these studies, however understanding the relative roles of reduced dimensionality and altered dielectric and steric environments on these processes is difficult. Motivated by these observations, we study peptide isomerization to investigate the effects of the air-water interface on a simple unimolecular reaction.\cite{srisailam2002influence,dentino1991role,smith1999alanine,brooks1993simulations} In this way we isolate the role of the altered solvation environment of the air-water interface on a biologically relevant reaction, free of complications from diffusion limitations and covalent bond breaking.

Computational and theoretical studies of reactions at interfaces are challenging. Simulations of interfaces require large systems, and often reactions of biological relevance are rare relative to the time-scales accessible to \textit{ab initio} methods\cite{Galib.2021,Niblett.2021,kattirtzi2017microscopic,serva2018combining,huggins2019biomolecular}. As a means of understanding some general effects of air-water interfaces on biochemical processes, we consider the effect of the air-water interface on the conformational changes of alanine dipeptide. 
Alanine dipeptide has long served as the model to explore peptide dynamics, as it possesses many of the essential features of  larger proteins including flexible torsional angles and the presence of CO, NH and CH$_3$ groups that mediate both hydrophilic and hydrophobic interactions.\cite{smith1999alanine,brooks1993simulations} Previous studies have demonstrated the importance of solvent in the reaction coordinate of isomerism\cite{drozdov2004role,Bolhuis.2000,Ma.2005,velez2009kinetics,Jung.2019}. \revision{Moreover, prior works on the mechanism of peptide isomerization at water-hexane\cite{Pohorille.1994}, water-gold\cite{Bellucci.2014} and water-bilayer\cite{pohorille1994interaction} interfaces have found significant discrepancies in both kinetic and thermodynamic observables near the interface. These inconsistencies along with the aforementioned characteristics emphasize the need of a systematic study of peptide isomerization at the air-water interface.}

In this work, we use molecular dynamics simulation to investigate the thermodynamics and kinetics of alanine dipeptide at the air-water interface. We observe an affinity of the peptide towards the interface, a phenomenon that has been observed in previous studies of other interfaces\cite{Bellucci.2014,Pohorille.1994}. We find that not all conformational states are equally stabilized, and particularly the relative likelihood of observing conformations intermediate between long lived dihedral states is decreased, suggesting a suppression of rate at the interface. An extension of transition path sampling, TPS+U\cite{Schile.2019} is used to probe the relative rate of isomerization between the metastable conformations $\alpha_L$ and $\beta$ as a function of the distance from the air-water interface, without prior mechanistic assumptions or the conservation of the mechanism near and away from the interface. This calculation confirms a 3-fold slow-down of isomerization near the air-water interface relative to its bulk value. We find an increase in favorable intramolecular peptide interactions that stabilize particular dihedral states at interface. This favorable intramolecular interaction is not observed for intermediate dihedral angles, from which we conclude that both the increased stability as well as the suppression of rate can be attributed to the availability of electrostatically favorable conformations that are inaccessible in bulk due to shielding from the solvent.

\begin{figure*}[!ht]
  \centering
    \includegraphics[trim={0cm 2cm 0cm 2cm},clip,width=\linewidth]{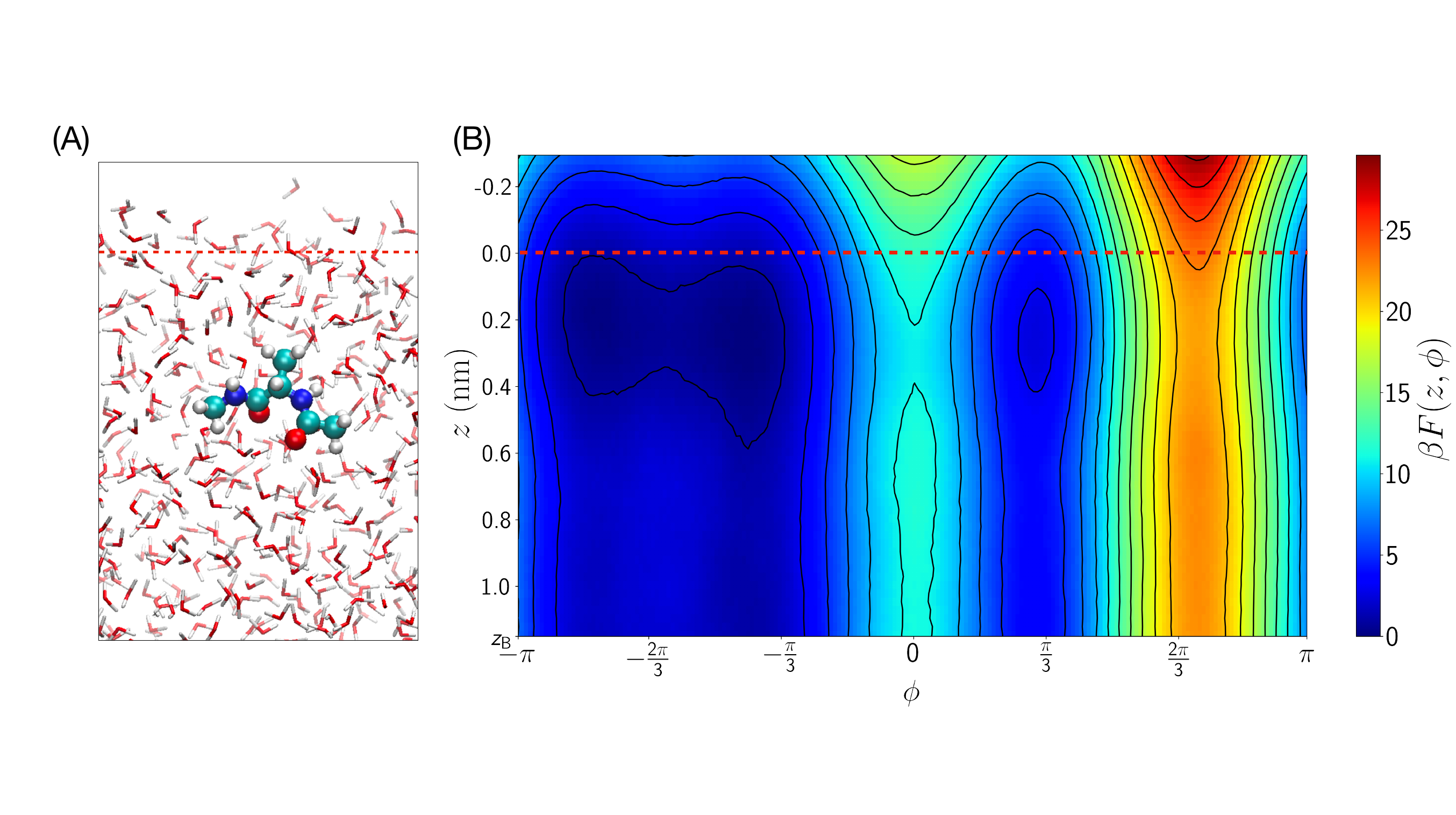}
    \caption{Stability of alanine dipeptide at the air-water interface. (A) Representative snapshot of the molecular dynamics simulation. (B) The free energy surface as a function of the dihedral angle $\phi$ and the distance from the Gibbs dividing surface $z$. Positive values of $z$ correspond to the bulk whereas negative values correspond to the vapor phase. The contours lines are spaced by 2 $k_{\mathrm{B}}T$.  The red dashed line in both figures denotes the Gibbs dividing surface.}
    \label{fig:2D_US}
\end{figure*}

{\bf Simulation details.}
The AMBER ff14SB forcefield\cite{maier2015ff14sb} parameters and TIP3P water model\cite{jorgensen1983comparison} are used to describe the molecular interactions of a solvated alanine dipeptide. The SETTLE algorithm\cite{miyamoto1992settle} is used to keep the water molecules rigid. All simulations are performed in OpenMM\cite{eastman2017openmm}. The system is comprised of 3000 water molecules and 1 alanine dipeptide and are placed in a periodic box of size 3x3x18 nm$^3$. A cutoff of 1.4 nm is used for computing short range interactions, and the Particle Mesh Ewald method\cite{essmann1995smooth} is employed to calculate long-range electrostatics. A Langevin integrator with a characteristic time-scale of 1 ps$^{-1}$ is used to maintain the temperature at 300 K for all simulations. A velocity Verlet discretization\cite{sivak2013using} is used to ensure time-reversibility.

For all calculations, the system is first equilibrated for 1 ns. After equilibration, the water slab is approximately 10.4 nm thick along the coordinate orthogonal to the air-water interface. The Gibbs dividing surface, defined as the plane where the density of water is half of that in bulk on average, is located 5.2 nm away from the center of mass of all water molecules. We employ a coordinate system where $z$ is defined as the distance from the Gibbs dividing surface, with positive values corresponding to liquid phase. A representative snapshot of the simulation is shown in Fig.~\ref{fig:2D_US}(A).

We investigate the transition between two lived conformations of alanine dipeptide, denoted the $\alpha_L$ to $\beta$ states. The C-N-C$_\alpha$-C torsional angle, denoted $\phi$, is used as the order parameter to separate these two conformations. Specifically, we employ the indicator functions 
\begin{equation} \label{eq:1}
h_{\alpha}(t) = \Theta[\phi(t)] \; \; \; \; \;
h_{\beta}(t) = \Theta[-\phi(t)]
\end{equation}
where $\phi$ is evaluated at time $t$ and $\Theta$ is the Heaviside function. Within each basin defined by the above indicator functions in the bulk liquid, both the $\alpha_L$ to $\beta$ states exhibit additional long lived conformations distinguished by other intramolecular coordinates.\cite{ferguson2011integrating,Vy} Thermodynamically, we must take care to ensure that sampling is ergodic within each basin.  Kinetically,  all rate calculations are started in the $\alpha_L$ state, for which there is a direct path to the $\beta$, free of long-lived intermediates.\cite{velez2009kinetics}

{\bf Thermodynamics near the air-water interface.}
Umbrella sampling is employed to investigate the thermodynamics of the two metastable isomers of alanine dipeptide near the air-water interface.\cite{frenkel2001understanding} Simulations are performed with a harmonic bias on $\phi$ and the $z$ component of the center of mass of alanine dipeptide of the form $ U(z,\phi) = K_\phi (\phi-\phi_0)^2/2 + K_z (z-z_0)^2/2$ with $K_\phi$ = 500 kJ mol$^{-1}$ rad$^{-2}$ and $K_z=200$ kJ mol$^{-1}$ nm$^{-2}$. 
A total of 20 windows are spaced 0.1 nm apart along $z$ using $z_0$ between -0.2 nm and 1.2 nm, and a total of 36 windows spaced 0.18 rad apart along $\phi$ using $\phi_0$ between $-\pi$ and $\pi$, equating to 720 windows in total. Each window is run for 5 ns and 1 ns of the simulation is discarded for equilibration. The Weighted Histogram Analysis Method \cite{kumar1992weighted} is used for estimating the free energies from the joint histograms in $\phi$ and $z$. 

The free energy is shown in Fig.~\ref{fig:2D_US}(B), defined from the joint probability,
\begin{equation}
F(z,\phi) = -\kB T \ln \langle \delta(z-z(0)) \delta (\phi-\phi(0)) \rangle
\end{equation}
expressed as an average, $\langle \dots \rangle$, over Dirac delta functions $\delta$ within the canonical ensemble at fixed temperature $T$ and $\kB$ is Boltzmann's constant. 
For all values of $z$, two stable conformations are observed around $\phi = \pi/3$ and  $\phi = - 2\pi/3$ with a barrier between them. These two regions correspond to the $\alpha_L$ and the $\beta$ phase, respectively. The free energy plateaus for $z>$ 1 nm away from the air-water interface, suggesting that the effect of the interface on the peptide stability decays within a few molecular layers. We define $z_\textrm{B} \, = \, 1.15 \, \textrm{nm}$ as the reference $z$ value for estimating bulk properties.  

\begin{figure}[t]
  \centering
    \includegraphics[trim={0.6cm 0.5cm 0.8cm 0.8cm},clip,width=\linewidth]{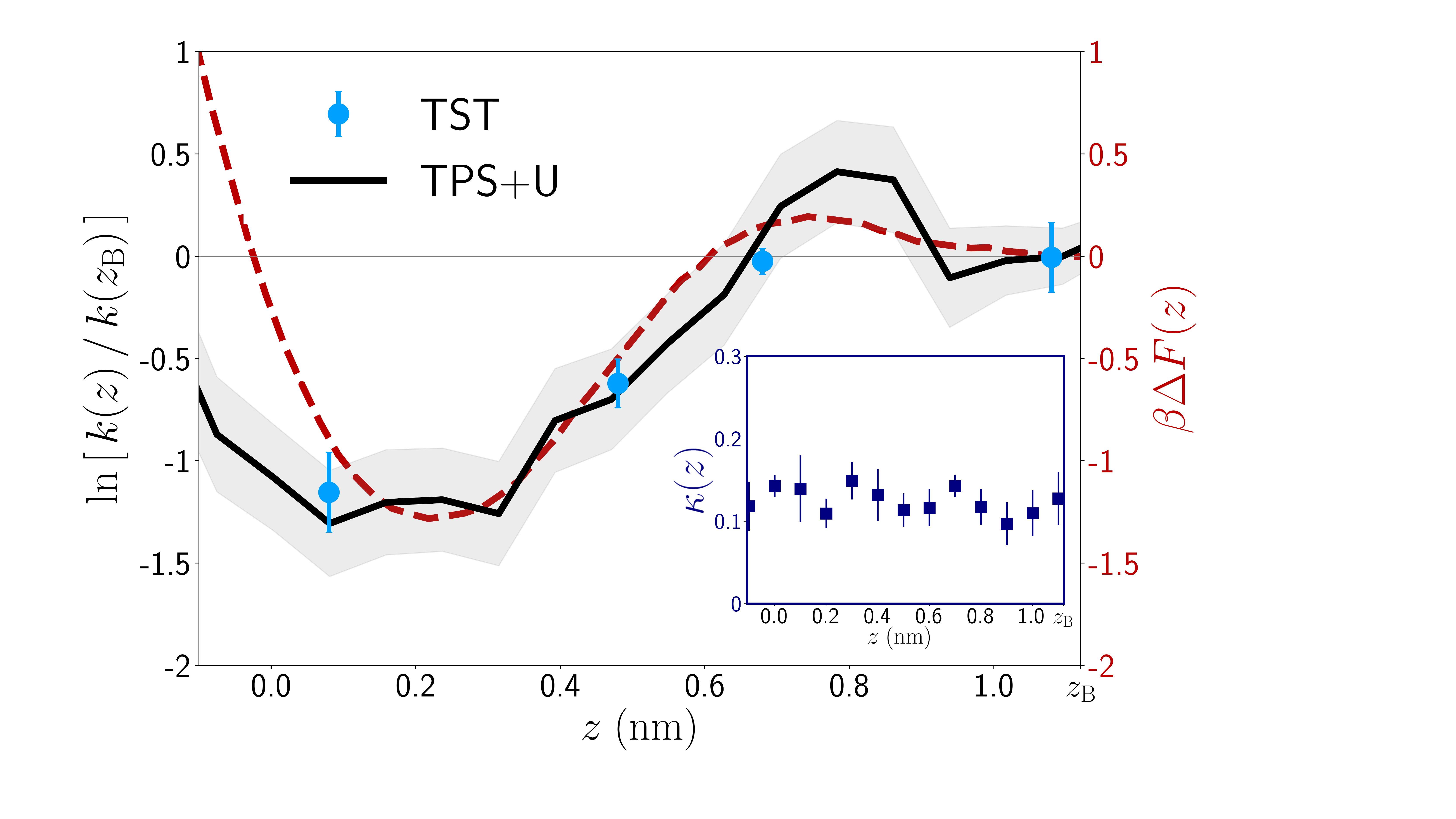}
    \caption{Kinetic and thermodynamic observables as a function of $z$. The red dashed line denotes the free energy relative to the bulk, $\Delta F(z) = F(z) - F(z_{\mathrm{B}})$. The black line and blue markers corresponds to the log ratio of the rate of $\alpha_L$ to $\beta$ transition as a function of $z$, computed using TPS+U and Transition State Theory respectively. The errorbars and the shading denote the standard error. \revision{Inset depicts the $z$ dependent transmission coefficient, $\kappa(z)$.} }
  \label{fig:rates}
\end{figure}

The $\beta$ conformation is more stable than $\alpha_L$ by about 2.5 $k_\mathrm{B} T$, in accordance with previous work with a similar forcefield.\cite{Vy} A considerable decrease in free energy is observed for moving either conformation to the water-vapor interface. This can be quantified by computing the free energy as a function of just $z$,
\begin{equation}
\label{Eq:Fz}
 F(z) = -k_{\mathrm{B}} T \ln \left[ \int_{-\pi}^{\pi} e^{-\beta F(z,\phi)} d\phi \right]
\end{equation}
by integrating over $\phi$. This relative free energy is shown in Fig. \ref{fig:rates} and reveals a negligible change from $z_\mathrm{B}$ to $z=$0.6 nm, below which the free energy starts to decrease. A minima of -1.4 $k_\mathrm{B} T$ relative to the bulk is observed around 0.2 nm away from the Gibbs dividing surface. This interfacial absorption is similar to studies of alanine dipeptide at gold-water\cite{Bellucci.2014} and water-hexane\cite{Pohorille.1994} interfaces. While the destabilization of the dipeptide in the vapor can be explained by the unfavorable process of desolvation, the increased in stability near the air-water interface reflects an interplay between energetic and entropic changes to the full system, explored below. 

The decrease in free energy near the interface is different for different $\phi$ values of the peptide. Both the $\alpha_L$ and the $\beta$ phases exhibit a 1.4 $k_\mathrm{B}T$ decrease in free energy near the interface, as computed by restricting Eq.~\ref{Eq:Fz} over the domains associated with the indicator functions in Eq.~\ref{eq:1}. We will refer to the $z$ dependent free energies for each state as $F_\alpha(z)$ and $F_\beta(z)$. As the change is the same for both conformers, the free energy difference between the two states, $F_\alpha(z)-F_\beta(z) $ is constant, with a value of $ 2.5\, \kB T$ that is independent of $z$ for the range of $z$ considered.
 However, the free energy of the dividing surface between the two phases, identified as $\phi = 0$, only decreases by 0.4 $k_\mathrm{B}T$. This inhomogeneous change in stability of different conformations results in an increase in the barrier height by about 1 $k_\mathrm{B}T$ near the interface, suggesting a suppression of fluctuations correlated with transitions between $\beta$ and $\alpha_L$ at the interface.

{\bf Kinetics near the air-water interface.}
The presence of the air-water interface breaks translational symmetry, and as a consequence the rate of isomerization between the $\alpha_L$ and $\beta$ conformations can in principle depend on the distance from the interface. In the limit that the diffusion of the dipeptide along $z$ is minimal during the characteristic transition path time,\cite{zwanzig1990rate,chekmarev2004long} we can associate an isomerization rate to each value of $z$. This separation of timescales exists for alanine dipeptide as the translation diffusion is small with a typical time to diffuse its diameter of nearly 1/2 ns\cite{rossky1979solvation}, permitting us to define and compute $k(z)$, the rate of isomerization as a continuous function of $z$.    

We define the spatially dependent rate constant using the general relationship between rate constants and path partition functions.\cite{dellago1998transition} Specifically, for observations times $t$ where the phenomenological rate constant is defined, the relative rate at $z$ and $z'$ is equal to
\begin{equation}
\frac{k(z)}{k(z')} = \frac{\Omega_{\alpha,\beta}(z)}{\Omega_{\alpha,\beta}(z')} e^{\beta  [F_\alpha(z)-F_\alpha(z')]}
\end{equation}
where $\Omega_{\alpha,\beta}(z)$ is the partition function associated with trajectories that start in $\alpha_L$ and end in $\beta$ at time $t$.This path partition function takes the form
\begin{equation}
\Omega_{\alpha,\beta}(z) = \int \mathcal{D}[\mathbf{X}(t)] P[\mathbf{X}(t)] h_{\alpha}(0) h_{\beta}(t) \delta [z-{z}(0)]
\end{equation}
where $P[\mathbf{X}(t)]$ is the probability to observe trajectory $\mathbf{X}(t)$, and $\delta [z-z(0)]$ constrains the transition to occur at $z$. This identification of the rate constant follows from the definition of a transition probability, or the standard time correlation function formalism.\cite{Chandler.1978,dellago2006transition}
TPS+U \cite{Schile.2019} provides a method to compute relative rates without prior mechanistic assumptions using this relation to path partition functions. The algorithm combines biasing potentials as in umbrella sampling with Transition Path Sampling\cite{Bolhuis.2002,dellago1998transition} to probe the relative rates, analogous to calculation of relative free energies.

To perform TPS+U, we generated initial reactive trajectories by propagating configurations at $\phi=0$ backward and forward in time for each value of $z$ studied. The length of the trajectories are kept fixed at 4 ps, which is sufficient for the dipeptide to diffuse over the barrier. The Shooting from the Top algorithm\cite{jung2017transition} is used to generate new decorrelated trajectories from old ones. This is done by randomly selecting configurations with $\phi$ in the range of $-0.3 \leq \phi \leq 0.3$ from the the old trajectory, drawing new velocities from the Maxwell-Boltzmann distribution, and propagating it forward and backward in time.

The number of shooting moves is varied according to $z$, as convergence is slow near the interface. Approximately 6000 shooting moves are used near the bulk, and around 12,000 moves are used for $z \leq 0.6$ nm. 
The acceptance rate varies between 0.2 to 0.3, resulting in 2000-4000 independent reactive trajectories. Harmonic biases of the same form used for the umbrella sampling but with $K_z$ between 100 and 200 nm$^{-2}$ used to constrain the trajectories at a given $z$. In total, 32 windows with minima spaced by 0.05 nm between $-0.2\leq z \leq 1.4$ nm are used together with histogram reweighting to compute the relative path partition function as a function $z$.

The results from TPS+U are shown in Fig. \ref{fig:rates}. The relative rate plateaus for $z>$ 0.8 nm to the bulk value, analogously as $F(z)$. Around $z=$0.6 nm, the rate decreases, and reaches a minima around $z=0.1$ nm before increasing towards its eventual value in the vapor. The magnitude of the suppression of the rate is approximately 2.5 times at $z=0$.  The suppression of the isomerization rate at the air-water interface is distinct from previous observations of smaller polar molecules which exhibit slight isomerization rate enhancements,\cite{benjamin1993isomerization} and more consistent with observed suppression of ion pair dissociation.\cite{dang2018rate,Niblett.2021,Galib.2021}

The observed rate suppression can be anticipated from the destabilization of conformations with $\phi= 0$ observed in $F(z,\phi)$. We can relate the rate suppression at the interface to this destabilization by comparing $k(z)$ with that approximated by transition state theory.\cite{nitzan2006chemical,Chandler.1978}  If we assume that the isomerization mechanism is invariant along $z$, and employ a diving surface as $\phi^{\ddag} = 0$, transition state theory gives us an estimate of the rate of $\alpha_L$ to  $\beta$ isomerization. Specifically, using $F(z,\phi)$, the transition state theory rate can be computed from
\begin{equation}\label{eq:2}
    k_{\mathrm{TST}}(z) = \frac{\langle | \dot \phi | \rangle}{2} \frac{e^{-\beta F(z,0)}}{\int_{0}^{\pi} e^{-\beta F(z,\phi)}d\phi}
\end{equation}
where $\langle | \dot \phi | \rangle$ corresponds to the average angular velocity, and the limits of the integral correspond to the $\alpha_L$ state. We find that the denominator in Eq.~\ref{eq:2} does not change with $z$. As a consequence the log ratio of rates $\ln [ k(z)/k(z_{\mathrm{B}})]$ is simply equal to the difference in barrier heights. This estimate is shown in Fig.~\ref{fig:rates}, and is consistent with TPS+U calculations, illustrating that the change in stability of conformations around $\phi=0$ is correlated with the fluctuations that result in isomerization. 

In order to determine the absolute rate, we have evaluated the rate in the bulk using the Bennett-Chandler method.\cite{Chandler.1978} The Bennett-Chandler method corrects the transition state theory approximation by defining a correlation function called the transmission coefficient, $\kappa(z,t)$,
\begin{equation}\label{eq:3}
\kappa (z,t) = \frac{\langle \dot \phi (0) \; \Theta[-\dot \phi (0)] \left[ h_\beta(t) - h_\beta(-t) \right] \rangle_{ \phi^\ddagger,z}}{\langle \dot \phi(0) \Theta[ -\dot \phi(0)] \rangle_{ \phi^\ddagger,z}} 
\end{equation}
where $\langle \dots \rangle_{\phi^{\ddag}, z}$ denotes an average over configurations on top of the dividing surface at a given value of $z$. The transmission coefficient decays quickly and reaches a plateau denoted $\kappa(z)$, and the value at the plateau determines the phenomenological rate. Using the transmission coefficient, the Bennett-Chandler rate is given by $k_{\mathrm{BC}}(z) = k_{\mathrm{TST}}(z) \kappa(z)$. To compute $\kappa(z_\mathrm{B})$, 30 configurations on top of the dividing surface are selected. Maxwell-Boltzmann velocities are drawn randomly, and the system is propagated forward and backwards in time for 2 ps.  A total of 2000 forward and backward trajectories are run for each configuration. The transmission coefficient plateaus within 0.3 ps seconds and is found to be small, 0.15 $\pm$ 0.03, yielding a rate in the bulk of 0.119 ns$^{-1}$. \revision{Additional calculations of $\kappa$ at different values of $z$ are unchanged from this bulk value as shown in Fig.~\ref{fig:rates}. Together with the observation of minimal changes in the torsional diffusion constant near the interface, these findings suggest that dynamical correlations between the solvent and $\phi$ are unaffected by the interface.}

{\bf Enhanced stability and rate suppression.}
As the relative suppression of rate near the air-water interface can be quantitatively captured by the $z$-dependence in the barrier height along the $\phi$ direction, the thermodynamics of the dipeptide is further investigated. We have analyzed the various components of the potential energy as a function of $z$ leveraging the pairwise additivity of the potential. 
To understand how these different interactions change near the interface, 10 ns simulations are performed with a harmonic potential along a range of $z$ values and two $\phi$ values denoting the $\alpha_L$ conformation and the dividing surface. 

\begin{figure}[t]
  \centering
    \includegraphics[width=1\linewidth]{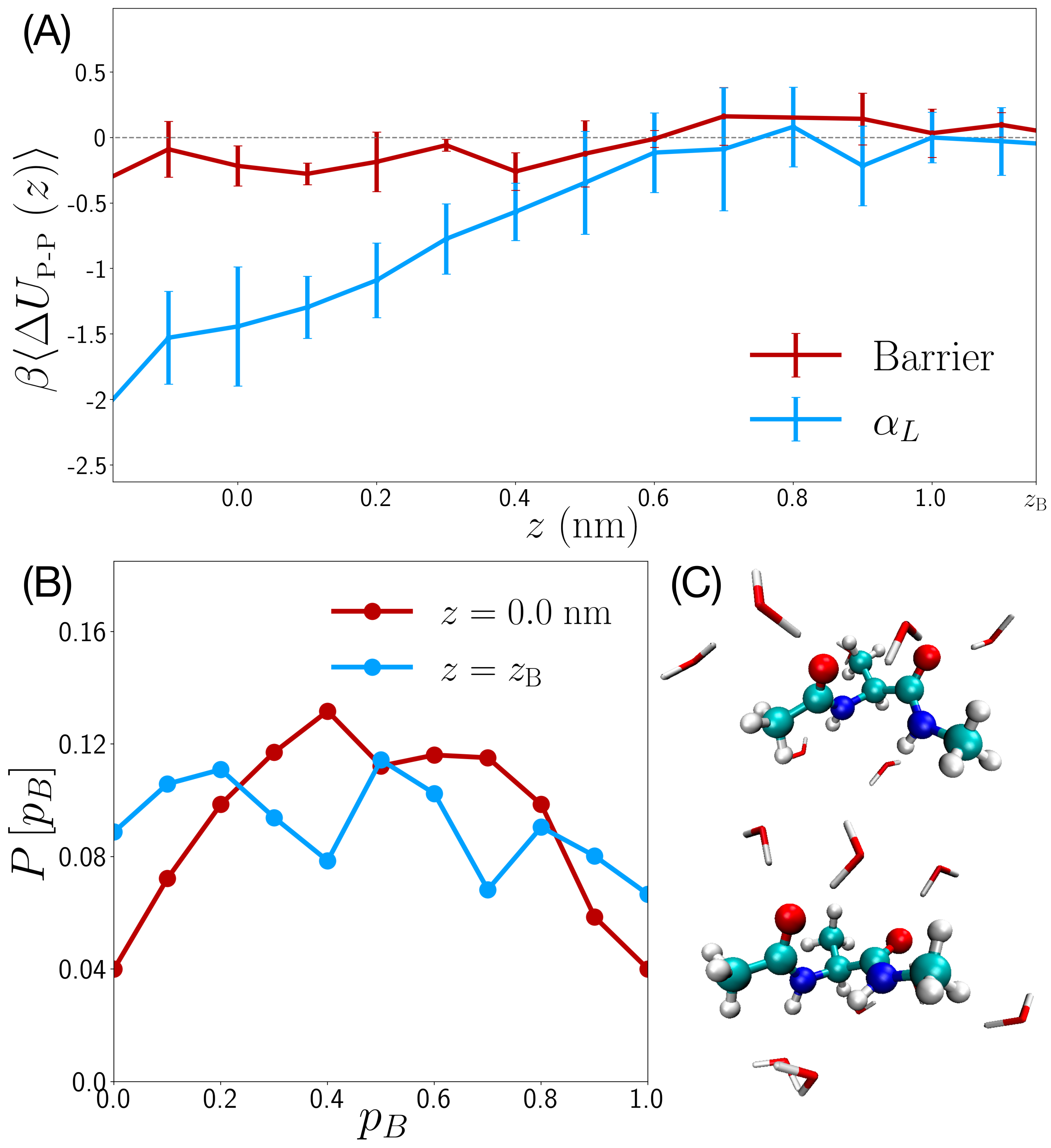}
    \caption{Importance of intramolecular peptide degress of freedom. (A) Change in the peptide-peptide potential energy as a function of $z$ for the barrier and the $\alpha_L$ conformation. Errorbars denote the standard error obtained from block averaging. (B) Commitor  probability  distribution  for  the  barrier  along $\phi$ ($-0.15\leq\phi\leq0.15$). (C) Visualization of the dipeptide along with the water in its first solvation shell at the GDS (top) and the bulk (bottom) is shown as reference.}
  \label{fig:committer}
\end{figure}

A prominent difference between the two states is observed in the average intramolecular potential of the dipeptide, $\langle \Delta U_{\mathrm{P-P}}(z)\rangle$ = $\langle U_{\mathrm{P-P}}(z)\rangle$ - $\langle U_{\mathrm{P-P}}(z_\mathrm{B})\rangle$, shown in Fig. \ref{fig:committer}(A). While the average intramolecular interactions are invariant along $z$ for conformations with $\phi=0$, these interactions decrease by around 2 $k_{\mathrm{B}}T$ for the $\alpha_L$ conformation at the interface relative to the bulk. A further decomposition of these interactions into bonded and nonbonded terms reveals that this enhanced stability arises exclusively from electrostatic interactions. \revision{These changes are not well approximated by a point dipole, due to their short ranged nature and the complexity of the charge distribution of alanine dipeptide.}
These results suggest that the decrease in water density at the interface allows the dipeptide to access certain electrostatically favorable conformations that are unfavorable in bulk. Analogous conformations are not accessible to the barrier state. In this way, as the dipeptide is desolvated, intramolecular interactions become unshielded by the water, stabilizing the metastable states $\alpha_L$ and $\beta$ and destabilizing configurations between them. 

These observations are in line with multiple studies that have underscored the role of water in the isomerization of alanine dipeptide.\cite{drozdov2004role,Bolhuis.2000,Ma.2005,velez2009kinetics,Jung.2019} Bolhuis et al.\cite{Bolhuis.2000} analyzed reactive pathways of  $\alpha_R$ to  $\beta$ isomerization of alanine dipeptide in both vacuum and solvated phases and found the dihedral angles to be a poor descriptor of the reaction coordinate. Ma and Dinner\cite{Ma.2005} suggested that information on the electrostatic interactions between the water and peptide was essential to capture the isomerization in explicit solvent. This was confirmed in a subsequent study by Velez-Vega et al\cite{velez2009kinetics}. More recently, Jung et al.\cite{Jung.2019} used neural networks with symmetry functions to encapsulate the effect of water required to describe the reaction coordinate. 

In order to reconcile the previously observed importance of solvent degrees of freedom with the success of transition state theory to predict the rate profile, we have analyzed the commitor probability distribution conditioned along $\phi=0$.\cite{dellago2006transition} The commitor, $p_\beta$, is a direct  indicator of the progression of a reaction. It is defined as the fraction of trajectories that go to $\beta$ before $\alpha_L$. By computing the commitor probability conditioned on a particular value of an order parameter, we can infer its correlation with the transition state ensemble.\cite{dellago2006transition,Geissler.1999} The commitor conditioned on $\phi=0$ at $z$ is given by,
\begin{equation} \label{eq:4}
p_\beta = \frac{\int \mathcal{D}\left[ \mathbf{X}(t) \right] P\left[ \mathbf{X}(t) \right]  \delta \left[\phi(0)\right]\delta \left[z-z(0)\right] h_\beta(t)}{\int \mathcal{D}\left[ \mathbf{X}(t) \right] P\left[ \mathbf{X}(t) \right]  \delta \left[\phi(0)\right]\delta \left[z-z(0)\right]}
\end{equation}
and calculated for 1000 configurations within a small window around $\phi=0$ that are selected from the TPS+U calculations. For each configuration we compute an estimate of $p_\beta$ by propagating 10 trajectories with random Maxwell-Boltzmann velocities, until they reach the $\alpha_L$ or $\beta$ state. The resultant distributions of $p_\beta$ from configurations drawn from the bulk and the interface are shown in Fig. \ref{fig:committer} (B).

A moderately flat distribution of commitment probabilities is observed in the bulk, suggesting that the transition occurs diffusively over the the thermodynamic barrier defined by $\phi=0$. This finding is consistent with previous studies that highlight the importance of solvent degrees of freedom in the reaction pathway of isomerization. By contrast, the distribution of $p_\beta$ at the interface resembles a Gaussian distribution peaked at 0.5. The probability of $p_\beta =0.5$ is nearly 3 times larger than that of $p_\beta =0$ and $p_\beta =1$, suggesting that that $\phi$ becomes a better reaction descriptor near the interface or that the effect of water becomes less significant to the isomerization pathway near the interface. 

The commitor calculations along with the potential decomposition suggest that isomerization is modulated by the reorganization of intramolecular dipeptide interactions. In bulk, these interactions are strongly dependent on the water structure around the peptide, making the water coordinates important in describing the reaction. Near the interface however, the role of the water structure diminishes, making $\phi$ an accurate descriptor of the reaction coordinate. However, due to the increased stability of the peptide near the interface, the energy fluctuations required to induce this reorganization become rarer, in turn suppressing the rate of peptide isomerization.

The physical and chemical properties of air-water interfaces can substantially alter the kinetics of biologically relevant reactions. Using isomerization of alanine dipeptide as a model, we have illuminated the interplay of interactions that arise at the air-water interface. Electrostatic interactions between the peptide dominate at the interface leading to a 4 times increase in stability. These interactions are strongly dependent on the conformation of the peptide, and we find the transition state of the $\alpha_L$ to $\beta$ isomerization to be unaffected by this stabilization. As a result, a three-fold suppression in the rate is observed at the air-water interface.

Our observations of selective affinity of peptides to the air-water interface as well as sluggish conformational dynamics they exhibit there both have their origins in the altered ability of the air-water interface to solvate polar solutes. These observations are consistent with similar mechanisms underlying the suppression of rates of other charge organization processes like hydrolysis\cite{Galib.2021} and ion pair dissociation.\cite{Niblett.2021} \revision{Direct comparison to the thermodynamics of adsorption and the role of energetic driving forces could be accomplished with spectroscopic titration measurements like those done with simple ions.}\cite{otten2012elucidating} Understanding the implication of these effects for more complex chemistry like peptide bond formation\cite{Griffith.2012} and acid dissociation\cite{shamay2007water,wang2009depth} is an important direction for future work.

\section*{Acknowledgements}
ANS and DTL were supported by NSF Grant CHE2102314.

\bibliography{ADP_interface}

\end{document}